# Nonlinear absorption of tetrathiafulvalene radical cation (TTF[+]) based charge transfer (CT) aggregates in PMMA

Barbara Federica Scremin[1*]

1- DSFTM –CNR Physical Sciences and Technology Department of the National Research Council of Italy

Correspondence to barbara.scremin@cnr.it

**Abstract**

Nolinear Absorption of CT TFF[+] aggregates in PMMA matrix was studied with resonant excitation around 800 nm in the femtoseconds regime with the Z-Scan technique with 100 Hz pulse repetition rate. The saturated absorption profiles were fitted with a prepared formula based on steady state rate equations for a homogeneous ensemble of two levels systems, thin samples and quasi monochromatic field without the usual assumption of low saturation regime. The lifetime of the CT state, with the use of this model was estimated in hundred of fs, of the same order of magnitude of the exciting pulse.

**Introduction**

TTF is studied even in supramolecular chemistry[1] for its π electron system and for the presence of three stable oxidation states that makes it attractive in the perspective of exploitation[1] as sensor, in logic gates and in redox switching. Here the nonlinear absorption properties of aggregates of the radical cation TTF[+] are studied in ultrafast regime. To our knowledge there are no examples of this kind of study in the literature. TTF[+] CT systems were included in PMMA matrix according to a previously described procedure[2] via solvent casting. The samples were 30 μm thick, and allowed to measure nonlinear optical properties with the Z-Scan technique[3], thus in a transmission configuration. Peculiar of the TTF[+] dimers and aggregates is the intense broad CT absorption band around 800 nm (fig. 1).

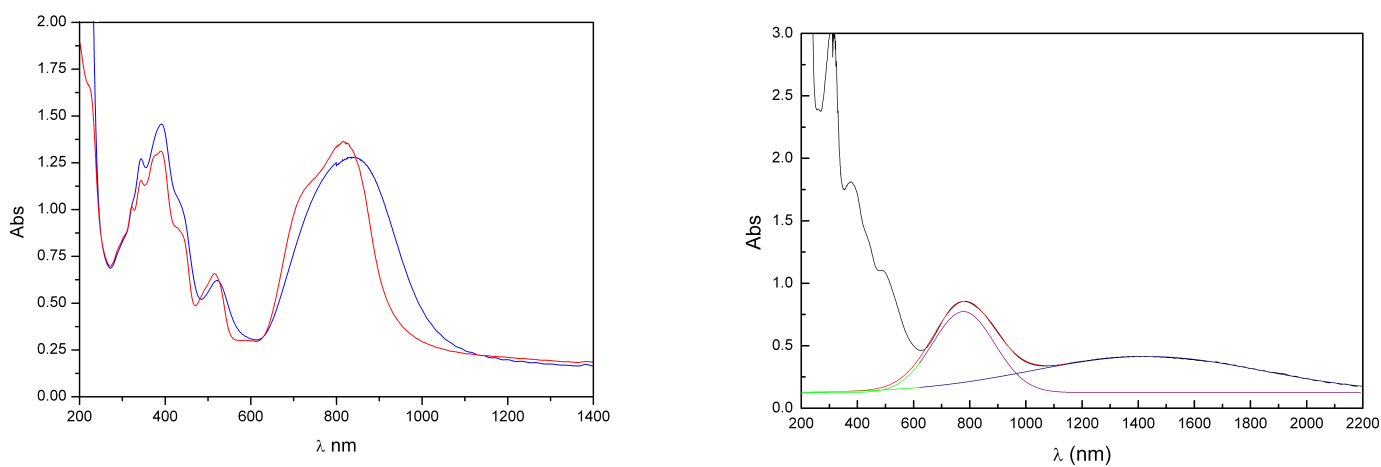

Fig. 1 Absorption spectra of TTFClO$_4$ included in PMMA thin films: on the left in blue the room temperature spectrum and in red the low temperature spectrum (15K) both under vacuum; on the right the room temperature spectrum of the film at equilibrium after some hours of air exposure.

TTF$^+$ salts form CT dimers in polar solvents[4] like DMSO but solutions are unstable[2] and crystals have high optical density preventing the measurements of their nonlinear optical properties in transmission configurations and for this reason a strategy of obtaining thins samples including in PMMA TTF$^+$ aggregates (from TTFClO$_4$) was developed in a previous work[2]. In fig. 1 absorption spectra of TTF$^+$ aggregates are shown for a thin film under vacuum and after some hours of air exposure: it was possible to notice that TTF$^+$ in PMMA formed up to a certain extent with air exposure TTF$^0$ (band around 300 nm) and TTF$^+$ TTF$^0$ based aggregates (band at 1400 nm). The interpretation was that the oxygen bound TTF$^{2+}$ and formed an S-oxide shifts the dismutation equilibrium of TTF$^+$ with the consequent formation of TTF$^+$ TTF$^0$.

This was in line with the previously studied dismutation reaction in solution[2] with a difference: in solutions, the dismutation reaction of TTF$^+$ fully evolved toward TTF$^0$ and TTF$^{2+}$, with the complete disappearance of the CT band of TTF$^+$ dimers and aggregates[2] while in PMMA at least the reaction reached an equilibrium in which the CT band was still present. The appearance of a new band around 1400 nm may be due to the formation of mixed valence aggregates, and may be representative of a lower energy charge transfer. In a first approximation, it was assumed to have excited around 800 nm aggregates of TTF$^+$ since this CT band was detected in dimers[1] and higher order aggregates[2,8] in PMMA.

**Experimental, Modelling and Results**

The Z-Scan technique was chosen to measure the absorptive nonlinearity in the open aperture scheme[3]. Briefly the thin sample was translated along a focused Gaussian beam experiencing in this way a continuous variation if the input intensity. A large area photodiode acquired the signal in the far field. The exciting system was constituted of a mode locked Titanium: Sapphire laser (Tsunami, Spectra Physics) pumped by an Ar$^+$ (Beam Lock, Spectra Physics) coupled to a regenerative amplifier with pulse stretcher and compressor (Spitfire, Spectra Physics) pumped by a solid-state intracavity doubled Nd: YLF laser (Merlin, Spectra Physics). The system delivered hundred fs pulses tuneable around 800 nm and repetition rate accordable between 1 KHz and 1 Hz. The Z-Scan setup consisted of a focusing lens (f=160 mm) aligned with a computer controlled (Lab View, National Instruments) motorized (MM2500 Newport) micrometric delay line (resolution 1 μm) on which the sample is mounted. The output signal was acquired in the far field with a large area silicon photodiode connected to a digital oscilloscope (TDS 520B 500 MHz Tektronix). The pulse temporal width was measured with an autocorrelator (Single Shot Autocorrelator SSA Positive Light), the pulse energy was measured with a pyroelectric detector (J3 Joulemeter Molectron) and the beam spatial profile was monitored with a CCD camera. The experimental result here reported (fig. 2) was obtained with 100 Hz of pulse repetition rate since from the tests it resulted not affected from cumulative or thermal effects on the basis of the nonlinearity response time[5]. The beam parameters were all experimentally determined allowing performing absolute measurements. The Z-Scan profile was fitted with an ad hoc analytical function formulated on the basis of a homogeneous ensemble of two levels system in steady state[6]:

$$\Delta N_{SS} = \frac{\Delta N_0}{1 + 2\tau \frac{\sigma I_{in}}{\hbar \omega}} \qquad (1)$$

Where $\Delta N_{SS}$ is the population difference in steady state $\Delta N_0$ the population difference at t=0, τ the population lifetime, σ the absorption cross section, ω the transition angular frequency, $I_{in}$ the input intensity. $I_{sat} = \hbar\omega/2\tau\sigma$ was defined as saturation intensity. The attenuation equation without any assumption on the input intensity versus the saturation intensity was:

$$\frac{dI(z)}{dz} = \alpha \frac{I(z)}{1+I(z)/I_{sat}} = \alpha_{tot}(I)I(z) \quad (2)$$

where α was the linear absorption coefficient. Integrating over a thickness dz and in the approximation that dz is sufficiently thin to allow the total absorption coefficient to remain constant

$$\int_{I(z=0)}^{I(z+dz)} \frac{dI(z)}{I(z)} = -\frac{\alpha}{1+I(z=0)/I_{sat}} \int_{z=0}^{z+dz} dz \quad (3)$$

an expression for the attenuated intensity for a sample of thickness dz was obtained:

$$I_{out}(dz) = I_{in}(z=0)\exp\left(\frac{-\alpha}{1+I_{in}(z=0)/I_{sat}}dz\right) \quad (4)$$

In our specific case we set dz=L since the thickness of the sample was 30 μm and had to be compared with the Rayleigh range $Z_0 = kw_0^2/2$ of the beam, where $w_0$ is the beam waist, the beam radius at the focus then $L<<Z_0$ and it is possible to work in the thin sample approximation. Substituting the expression for the intensity distribution for a pulsed Gaussian beam

$$I_{in}(t,r,Z) = I_{in}(t,0,Z)\exp\left(\frac{-2r^2}{w^2(Z)}\right) \quad (5)$$

and integrated on the Gaussian intensity distribution gave the output power in a position $Z_j$ along the focused beam

$$P_{out}(t,Z_j) = \int_0^{+\infty} I_{out}(t,r,Z_j) 2\pi r dr$$

$$= \frac{2P_{in}(t,Z_j)}{\pi w^2(Z_j)} 2\pi \int_0^{+\infty} \exp\left(\frac{-2r^2}{w^2(Z_j)}\right) \exp\left(\frac{\alpha L}{1+I_{sat}^{-1}I_{in}(t,r,Z_j)}\right) r dr \quad (6)$$

substituting the explicit expression (5) for the input intensity in the expression for the output power (6) we obtain a transcendent form which is not integrable analytically

$$P_{out}(t,\underset{\sim}{Z}_j) = \frac{4P_{in}(t,Z_j)}{w^2(\underset{\sim}{Z}_j)} \int_0^{+\infty} \exp\left(\frac{-2r^2}{w^2(\underset{\sim}{Z}_j)}\right) \cdot$$

$$\cdot \exp\left[\frac{-\alpha L}{1 + I_{sat}^{-1} \dfrac{2P_{in}(t,\underset{\sim}{Z}_j)}{\pi w^2(\underset{\sim}{Z}_j) \exp\left(\dfrac{-2r^2}{w^2(\underset{\sim}{Z}_j)}\right)}}\right] r\, dr \qquad (7)$$

Approximating the expression using the peak value of the radial distribution and a weight factor $c_r$ it was possible to make the second exponential factor independent from the integral sign and obtaining in this way the integral of a Gaussian distribution which allowed to obtain an estimation of the output power from an analytical integration

$$P_{out}(t,Z_j) \approx P_{in}(t,Z_j)\exp\left(\frac{-\alpha L}{1+I_{sat}^{-1}\dfrac{2P_{in}(t,Z_j)}{\pi w^2(Z_j)}c_r}\right) \qquad (8)$$

The weight factor $c_r$ was numerically evaluated at about 0.5 for an error on the power estimation of 2%. Analogously, integrating the instant power along the pulse temporal profile we calculate the nonlinear transmittance, normalized for the linear one with open aperture configuration:

$$T_N(Z_j) = \frac{T_{NL}(Z_j)_{open}}{T_L(Z_j)_{open}} = \\
= \frac{\int_{-\infty}^{+\infty} P_{out_{NL}}(t,Z_j)_{open}\, dt}{\int_{-\infty}^{+\infty} P_{out_L}(t,Z_j)_{open}\, dt} \qquad (9)$$

where the linear transmittance is obtained when the sample is far from the focus, the intensity is low and it behaves in a linear way ($Z_j \gg Z_0$, $Z_j \ll Z_0$). Introducing the pulse temporal profile from the autocorrelation measurements[7] into the expression for the input power and fixing to zero time the pulse peak value we obtain:

$$P_{in}(t) = P_{in}(t_0 = 0)\exp\left(\frac{-4\ln 2(t-t_0)^2}{\Delta\tau_{FWHM}^2}\right) \qquad (10)$$

Where

$$P_{in}(t_0 = 0) = \left(\frac{\varepsilon_{n/pulse}}{\Delta\tau_{FWHM}} 2\sqrt{\frac{\ln 2}{\pi}}\right) \qquad (11)$$

then substituting $P_{in}(t, Z_j)$ into $P_{out}(t, Z_j)$ and integrating over time to obtain the nonlinear transmittance another transcendent expression is obtained

$$T_{NL}(Z_j)_{open} = \frac{\varepsilon_{n/pulse}}{\Delta\tau_{FWHM}} 2\sqrt{\frac{\ln 2}{\pi}} \int_{-\infty}^{+\infty} \exp\left(\frac{-4\ln 2(t-t_0)}{\Delta\tau_{FWHM}}\right) \cdot$$

$$\cdot \exp\left[\frac{-\alpha L}{1 + \frac{1}{I_{sat}} \cdot \frac{2\left(\frac{\varepsilon_{n/pulse}}{\Delta\tau_{FWHM}} 2\sqrt{\frac{\ln 2}{\pi}}\right) \exp\left(\frac{-4\ln 2(t-t_0)^2}{\Delta\tau_{FWHM}}\right)}{\pi w^2(Z_j)} c_r}\right] dt \qquad (12)$$

again for the time integration introducing another weight factor $c_t$ evaluated at about 0.7 for an error of about 2% on the transmittance estimation we obtained

$$T_{NL}(Z_j) \approx \varepsilon_{n/pulse} \exp\left(\frac{-\alpha L}{1 + \frac{1}{I_{sat}} \cdot \frac{2\left(\frac{\varepsilon_{n/pulse}}{\Delta\tau_{FWHM}} 2\sqrt{\frac{\ln 2}{\pi}}\right) c_t}{\pi w^2(Z_j)} c_r}\right) \qquad (13)$$

The linear transmittance is easily calculated from the transmitted power in linear regime and was

$$T_L(Z_j) = \int_{-\infty}^{+\infty} dt \int_0^{+\infty} I_{out}(t,r,Z_j) 2\pi r\, dr =$$
$$= \varepsilon_{n/pulse} \cdot e^{-\alpha L} \qquad (14)$$

The normalized transmittance, with the $Z_j$ dependence became

$$T_N(Z_j) = \exp\left[\alpha L\left(1 - \cfrac{1}{1 + \cfrac{1}{I_{sat}} \cdot \cfrac{2\left(\cfrac{\varepsilon_{n/pulse}}{\Delta\tau_{FWHM}}\right)2\sqrt{\cfrac{\ln 2}{\pi}}c_t}{\pi w_0^2\left(1 + \cfrac{(Z_j - Z_c)^2}{Z_0^2}\right)}c_r}\right)\right] \quad (15)$$

Being $w_0$ the beam waist $w_0^2 = Z_0/\pi\lambda$, $Z_c$ the coordinate of the focus, the nonlinear parameter could then be obtained from the fitting of the experimental profiles introducing the experimental parameters as evaluated from direct measurements. The experimental parameters previously determined are resumed in the following table 1

Table 1. The experimentally determined values used in the data fitting

| λ [nm] | $\Delta\tau_{FWHM}$[sec] | $Z_0$ [cm] | L [μm] | σ[cm$^2$/dimer] | αL |
|---|---|---|---|---|---|
| 792 | 125·10$^{-15}$ | 0.159 | 30 | 1.10·10$^{-18}$ | 1.47 |

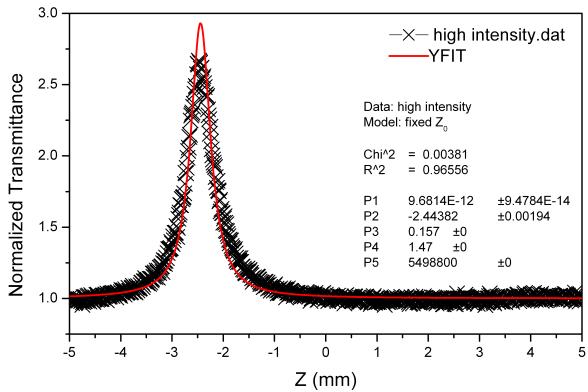 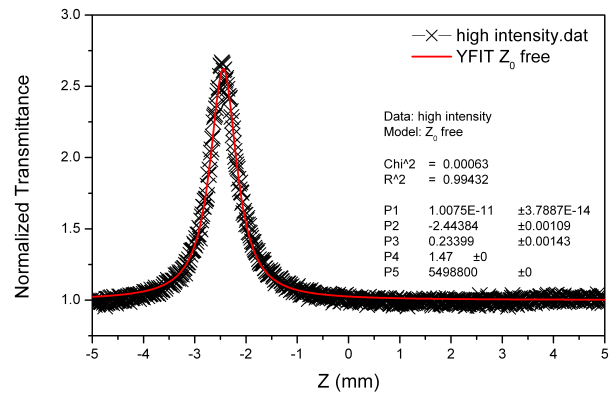

Figure 2. Fitted Z-Scan profile with fitting parameters: P1=1/$I_{sat}$, P2=$Z_c$, P3=$Z_0$, P4= $\alpha_L$L, ($\varepsilon_n$/$_{pulse}$)/$\Delta\tau_{FWHM}$. On the left side, the fitting was performed using all the experimentally determined parameters, on the right the diffraction length was left free.

In fig. 2 results from the fitting obtained inserting the prepared formula in Origin 6 are presented: the profile with all the parameters fixed overestimated the maximum, then the diffraction length of the beam was set free to vary and the Z-scan profile was reproduced more accurately. The parameter that should come out from the fitting was the saturation intensity that allowed to calculate the lifetime of the CT state. Results are shown in fig.3.

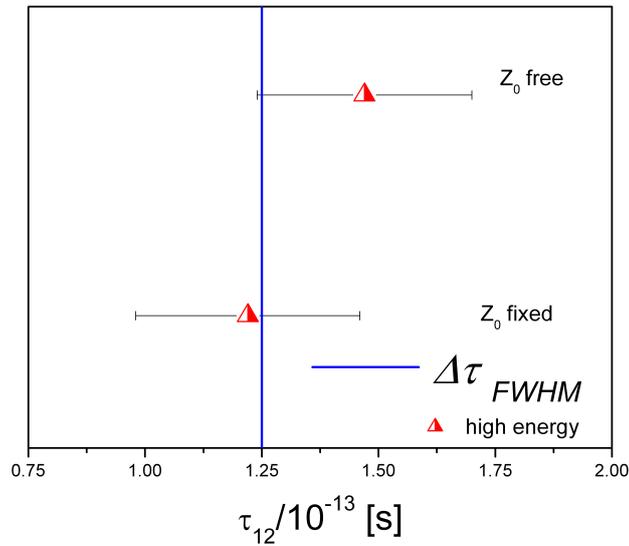

Figure 3. Lifetime of the CT state estimated from the fitting of the Z-scan data, with fixed and free $Z_0$ according to fig. 2. In blue the pulse length was evidenced.

**Discussion**

According to the obtained results it was possible to notice that under the errors the lifetime of the excited state was of the same order of magnitude of the exciting pulse width, which made weak the assumption of a steady state used to model the ensemble of two levels system (eq. 1) and to obtain the absorption coefficient (eq. 2). In 2006 a paper Gu et al. [9] developed a theoretical model for open aperture Z-scan based on the application of the Adomian decomposition method[10], to avoid the initial assumption of low saturation regime used in the original formulation[3], but they used steady state models for the absorption coefficient and calculated from the final data fitting the saturation intensity, which is in reality a function of the excited state lifetime. In the present work the assumption of low saturation regime ( input intensity << saturation intensity) was not necessary, the approach uses approximations to evaluate transcendent forms in integrals and did not use series expansions as in ref. 9; the final formula was quite simple (once the experimental parameters values are introduced), used experimentally determined values for the beam parameters and the sample and could be easily introduced in common data analysis programs to perform the fitting of the experimental Z-scan traces provided that a steady state assumption might subsist. To the author knowledge a model for the absorption coefficient without the assumption of steady state currently is not available in the literature.

**Conclusions**

The lifetime of the CT state for $TTF^+$ aggregates was estimated from Z-Scan experimental profiles. The profile fitting was performed with an ad hoc formulated fitting function based on a homogeneous ensemble of two levels systems in steady state. In this way no a priori hypothesis on the magnitude of the saturation intensity respect to the input intensity had to be assumed, since the conventional formulation is based in the low saturation regime hypothesis and had bad performances in fitting the experimental data, since in the femtosecond regime the intensity is usually quite high and nonlinearity may have contribution from orders higher than the third. Therefore the obtained lifetime, of the same order of magnitude of the pulse may give rise to questions regarding the adequacy of a steady state hypothesis. Moreover pure $TTF^+$ based systems resulted very difficult to handle: in air they evolved up to an equilibrium among $TTF^+$, $TTF^0$, $TTF^{2+}$. Around 800 nm it was excited a CT transition of $TTF^+$ aggregates/dimers around 800 nm, in such a case it may be adequate the modelling with a two level system. As discussed the lifetime of the excited state extracted from the data fitting was of the same order of magnitude of the exciting pulse duration and the steady state assumption for the calculation of the absorption coefficient could be not robust enough. In the literature the parameter extracted from the fittings is the nonlinear absorption, the fitting performs well, but the timescales are masked since the lifetimes usually are not calculated, then the steady state regime assumption under all the available fitting formulas is not discussed.

**Acknowledgements**

BFS acknowledge Prof. Renato Bozio who inspired the work during BFS PhD period at the University of Padova (1997-2000), Prof. Camilla Ferrante for introducing BFS to the femtosecond spectroscopy and her participation together with Prof. Danilo Pedron in setting the Z-scan experiment.